\documentstyle[twoside,fleqn,espcrc2,epsf]{article}

\title{
Equation of state for pure SU(3) gauge theory
on anisotropic lattices\thanks{Talk presented by Y.~Namekawa}
}

\author{CP-PACS Collaboration : 
        S.~Aoki\rlap,\address{Institute of Physics, University of
        Tsukuba, Tsukuba, Ibaraki 305-8571, Japan}
        R.~Burkhalter\rlap,$^{\rm a,}$\address{Center for 
        Computational Physics, University of Tsukuba, Tsukuba, 
        Ibaraki 305-8577, Japan}
        S.~Ejiri\rlap,$^{\rm b}$
        M.~Fukugita\rlap,\address{Institute for Cosmic Ray Research,
        University of Tokyo, Kashiwa 277-8582, Japan}
        S.~Hashimoto\rlap,\address{High Energy Accelerator Research 
        Organization(KEK), Tsukuba, Ibaraki 305-0801, Japan}
        N.~Ishizuka\rlap,$^{\rm a,b}$
        Y.~Iwasaki\rlap,$^{\rm a,b}$
        K.~Kanaya\rlap,$^{\rm a}$
        T.~Kaneko\rlap,$^{\rm d}$
        Y.~Kuramashi\rlap,$^{\rm d}$
	V.~Lesk\rlap,$^{\rm b}$
	Y.~Namekawa\rlap,$^{\rm a}$
	M.~Okamoto\rlap,$^{\rm b}$
        M.~Okawa\rlap,$^{\rm d}$
	Y.~Taniguchi\rlap,$^{\rm a}$
        A.~Ukawa\rlap,$^{\rm a,b}$ and
        T.~Yoshi\'e$^{\rm a,b}$ }

\begin{document}

\begin{abstract}
We present results for the equation of state for pure
SU(3) gauge theory obtained on anisotropic lattices
with the anisotropy $\xi \equiv a_s/a_t = 2$.
The pressure and energy density are calculated 
on $N_t / \xi = 4, 5$ and $6$ lattices
with the integral method.
They are found to 
satisfy the leading $1/N_t^2$ scaling from our coarsest lattice 
$N_t/\xi=4$. 
This enables us to carry out well controlled continuum extrapolations.
We find that the pressure and energy density
agree with those obtained using the isotropic plaquette action,
but have smaller and more reliable errors.
\end{abstract}

\maketitle
\thispagestyle{empty}

%%%%%%%%%%%%%%%%%%%%%%%%%%%%%%%%%%%%%%%%
\section{Introduction}

The study of lattice QCD at finite temperatures is important 
to clarify the dynamics of the quark gluon plasma \cite{ejiri}.
In order to extract predictions for the real world from 
lattice data, we have to extrapolate 
them to the continuum limit.
However
the continuum limit of the equation of state (EOS) has not been
obtained in full QCD because of large lattice spacing dependence
\cite{milc,bielefeld,cppacs_ft}.

We have proposed using anisotropic lattices to solve this problem.
As a first test, 
we have applied them for calculations of 
the EOS in pure SU(3) gauge theory
and studied their efficiency \cite{our_paper}.
We report a summary of our results in this article.

\section{Simulation}
\label{sec:simulation}

We employ the plaquette action given by 
\begin{equation}
S
= \beta
  ( 
    \frac{1}{\xi_0} \sum_{n,ij} \left(1 - P_{ij}(n)\right)
    + \xi_0 \sum_{n,i} \left(1 - P_{i4}(n)\right)
  )
\label{lat-gauge-aniso}
\end{equation}
where $\xi_0$ is the bare anisotropy and $\beta=6/g_0^2$.
We use the values of the renormalization factor 
$\eta(\beta,\xi) \equiv \xi / \xi_{0}(\beta,\xi)$ 
obtained by Klassen \cite{klassen}, 
where $\xi \equiv a_s/a_t$.

Our study in the Stephan-Boltzmann limit shows that
$\xi=2$ is the optimal choice to reduce discretization errors
in the EOS \cite{our_paper}.
Therefore we use
$\xi = 2$ anisotropic 
lattices with sizes $N_s^3\times N_t = 16^{3} \times 8$, 
$20^{3} \times 10$ and $24^{3} \times 12$. 
For $N_t=8$, we make additional runs on $12^3\times8$ and $24^3\times8$ 
lattices to examine finite size effects. 
The zero-temperature runs are made on $N_s^3\times \xi N_s$ lattices 
with $\xi=2$.
After thermalization, 
we perform 20,000 to 100,000 iterations on finite-temperature lattices
and 5,000 to 25,000 iterations on zero-temperature lattices.

\section{Scale and critical temperature}

We determine the physical scale of our lattices 
from the string tension.
The string tension $\sigma$ is extracted 
from the static potential $V(R)$
at zero temperature
assuming a form \cite{michael},
\begin{equation}
 V(R) = V_{0} + \sigma R - e \frac{1}{R}
                       + l \left( \frac{1}{R}
                       - \left[\frac{1}{R}\right]  \right),
\label{pot-fit}
\end{equation}
where $[1 / R]$ is the lattice Coulomb 
term from one gluon exchange.
Then we fit the string tension in $\beta$
using an ansatz proposed by Allton \cite{allton},
\begin{equation}
 a_s \sqrt{\sigma} 
 = f(\beta) \, (1 + c_{2}\hat{a}(\beta)^{2} 
                    + c_{4}\hat{a}(\beta)^{4}) / c_{0},
\label{allton_fitting_eq}
\end{equation}
where $f(\beta)$ is the two-loop scaling function of 
SU(3) gauge theory and  $\hat{a}(\beta) \equiv f(\beta)/f(\beta=6.0)$.

We define the critical coupling $\beta_{c}(N_t,N_s)$ from \\
the peak location of the susceptibility 
for a Z(3)-rotated Polyakov loop. 
We extrapolate $\beta_{c}(N_t,N_s)$
to the thermodynamic limit 
assuming a finite-size scaling law,
\begin{equation}
 \beta_{c}(N_t,N_s) = \beta_{c}(N_t,\infty) 
       - h \left( N_t / (\xi N_s) \right)^{3}.
\end{equation}
From
$\beta_{c}$ on anisotropic $12^3\times8$, $16^3\times8$ 
and $24^3\times8$ lattices with $\xi=2$, we find $h = 0.031(16)$ 
for $N_t/\xi=4$.
We adopt this value for all $N_t$.
The critical temperature in units of the string tension is given by
$T_c / \sqrt{\sigma} = 
 \xi / N_t a_s \sqrt{\sigma}\left(\beta_c(N_t,\infty)\right)$.

We extrapolate results for $F = T_{c} / \sqrt{\sigma}$ 
to the continuum limit 
with the leading scaling form,
\begin{equation}
\left. F \right|_{N_t} 
= \left. F\right|_{\rm continuum} + c_F / N_t^2.
\label{continuum-extrapolation}
\end{equation}
In the continuum limit, we obtain 
$ T_{c} / \sqrt{\sigma} = 0.635(10)$
from the $\xi=2$ plaquette action.

\begin{figure}[tb]
\vspace{-5mm}
\begin{center}
\leavevmode
\epsfxsize=7.5cm
\epsfbox{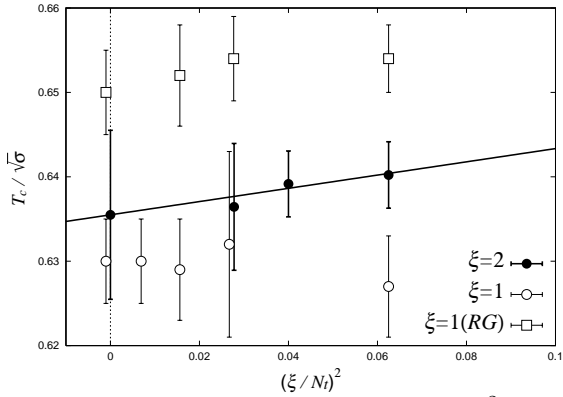}
\end{center}
\vspace{-14mm}
\caption{
$T_c/\protect\sqrt{\sigma}$ as a function of $(\xi/N_t)^2$. 
}
\label{fig:Tc}
\vspace{-15pt}
\end{figure}

Fig.~\ref{fig:Tc} shows our results of $T_{c} / \sqrt{\sigma}$.
In Fig.~\ref{fig:Tc}, we also plot the results obtained on isotropic 
lattices using the plaquette action \cite{beinlich2} and the RG-improved 
action \cite{tsukuba,okamoto}. 
Our value of $T_c/\sqrt{\sigma}$ in the continuum limit is consistent 
with that from the $\xi=1$ plaquette action
within the error of about 2\%,
but 
slightly smaller than that from the $\xi=1$ RG-improved action.
This may be caused by the difference in the details of 
the potential fit.

\section{Equation of state}

We calculate the pressure $p$
and energy density $\epsilon$
using the integral method\cite{engels4},
\begin{eqnarray}
\left. \frac{p}{T^{4}} \right|^{\beta}_{\beta_{0}} 
&=& \int_{\beta_{0}}^{\beta} d \beta^{\prime}
\Delta S
, \\
\frac{\epsilon - 3 p}{T^{4}} 
&=& - a_s \left. \frac{\partial \beta}
                      {\partial a_s} \right|_{\xi=2} \Delta S,
\label{pressure}
\end{eqnarray}
where
\begin{equation}
\Delta S 
\equiv  \left(\frac{N_t^{4}}{\xi^{3}}\right)
                              \frac{1}{N_s^{3}N_t}
   \left. \frac{\partial \log Z}{\partial \beta} \right|_{\xi=2}
   - (T=0)
.
\label{delta_S}
\end{equation}
The beta function
$\left. \partial \beta / \partial a_s \right|_{\xi=2}$
is determined through the string tension
parametrized by Eq.~(\ref{allton_fitting_eq}).
We extrapolate the EOS to the continuum limit 
with Eq.~(\ref{continuum-extrapolation}).

\begin{figure}[tb]
\vspace{-5mm}
\begin{center}
\leavevmode
\epsfxsize=7.5cm
\epsfbox{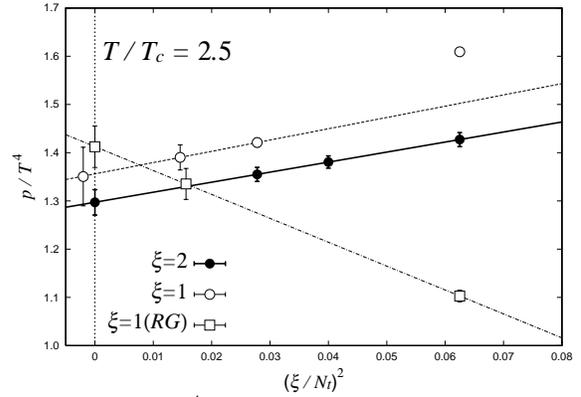}
\end{center}
\vspace{-14mm}
\caption{
$p/T^4$ at $T=2.5T_{c}$.
}
\label{fig-pressure-scaling}
\vspace{-15pt}
\end{figure}

\begin{figure}[tb]
\vspace{-5mm}
\begin{center}
\leavevmode
\epsfxsize=7.5cm
\epsfbox{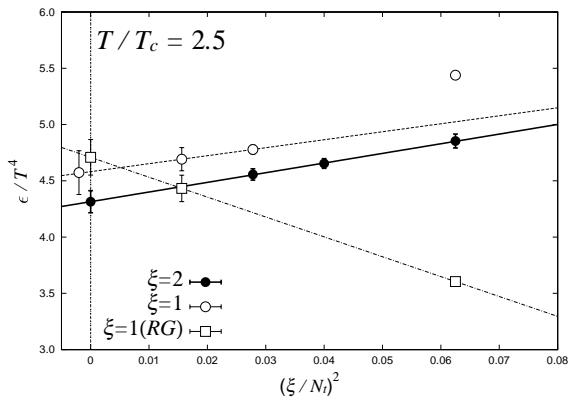}
\end{center}
\vspace{-14mm}
\caption{
$\epsilon/T^4$ at $T=2.5T_{c}$.
}
\label{fig-energy-scaling}
\vspace{-15pt}
\end{figure}

Figs.~\ref{fig-pressure-scaling} and \ref{fig-energy-scaling}
show the pressure and energy density at $T/T_c=2.5$, 
as a function of $(\xi/N_t)^2$ (filled circles). 
For comparison, results from isotropic lattices using the 
plaquette action \cite{boyd} (open circles) 
and the RG-improved action \cite{okamoto} (open squares) 
are also plotted. 

The advantage of anisotropic lattices is 
clearly seen in
Figs.~\ref{fig-pressure-scaling} and \ref{fig-energy-scaling}.
On the coarsest lattice $N_t/\xi=4$, the cutoff errors 
at $\xi=2$ are much smaller than those at $\xi=1$ 
with the same plaquette action. 
Comparing the computational cost to achieve comparable systematic 
and statistical errors on isotropic and $\xi=2$ anisotropic lattices,
we find that 
the anisotropic calculation has a factor of approximately 5 gain.
Furthermore, since the $\xi=2$ data satisfy
the $1/N_t^2$ scaling from
$N_t/\xi=4$ (the right-most point), 
we can reliably perform an extrapolation 
to the continuum limit using three data points. 
Results at other temperatures are similar.

\begin{figure}[tb]
\vspace{-5mm}
\begin{center}
\leavevmode
\epsfxsize=7.5cm
\epsfbox{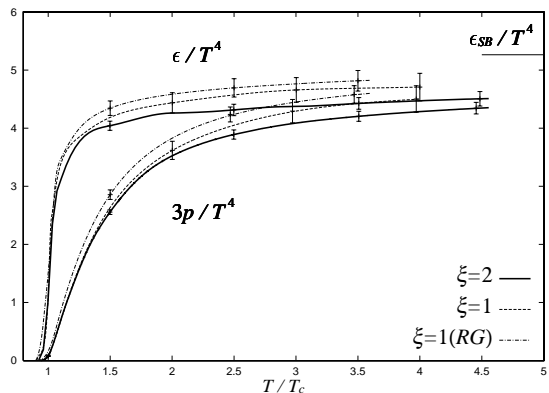}
\end{center}
\vspace{-14mm}
\caption{
$p/T^4$ and $\epsilon/T^4$ as functions of $T/T_c$. 
}
\label{fig-pressure_energy-continuum}
\vspace{-15pt}
\end{figure}

Our results for the EOS in the continuum limit are
shown in Fig.~\ref{fig-pressure_energy-continuum}.
The EOS from $\xi=2$ is
consistent with that 
from the $\xi=1$ plaquette action
within the error of about 5\%.
Notice that our $\xi=2$ results have smaller 
and more reliable errors of 2--3\%. 
On the other hand, 
the EOS from the $\xi=1$ RG-improved action
is larger by 7--10\% (about 2$\sigma$).
than that from our $\xi=2$ results in the continuum limit.
A possible origin of this discrepancy is the use of 
the $N_t/\xi=4$ data of the RG-improved action, which show a large 
(about 20\%) deviation from the continuum limit.

\section{Conclusion}
\label{sec:summary}

We have studied the continuum limit of the EOS 
in pure SU(3) gauge theory on $\xi = 2$ anisotropic lattices,
using the plaquette action. 
Anisotropic lattices are shown to be efficient 
in reducing the cutoff dependence of the EOS.
As a result, we can perform continuum extrapolations
from our coarsest lattice $N_t/\xi = 4$.
The EOS in the continuum limit agrees with that obtained 
on isotropic lattices using the same action,
but has much smaller and better controlled errors. 

Anisotropic lattices may help us extract
the continuum limit of the EOS
in full QCD.

\vspace{4mm}

This work is supported in part by Grants-in-Aid of the Ministry of Education 
(Nos.~10640246, % ukawa
10640248, % kanaya
11640250, % ruedi
11640294, % okawa
12014202, % s-aoki
12304011, % iwasaki
12640253, % s-aoki
12740133, % ishizuka
13640260  % kanaya
). 
M. Okamoto is JSPS Research Fellows. 
VL is supported by the Research for Future Program of JSPS
(No. JSPS-RFTF 97P01102).
Simulations were performed on CP-PACS 
at the Center for Computational Physics, University of Tsukuba.

%%%%%%%%%%%%%%%%%%%%%%%%%%%%%%%%%%%%%%%%%%%%%%%%%%%%%%%%%%%%%

%%%%%%%%%%%%%%%%%%%%%%%%%%%%%%%%%%%%%%%%%%%%%%%%%%%%%%%%%%%%%

\end{document}